\documentclass[aps, prc, floatfix, nofootinbib, superscriptaddress, twocolumn]{revtex4-1}

\usepackage{latexsym}
\usepackage{amsmath}
\usepackage{amssymb}
\usepackage{amsfonts}

\usepackage{color}

\usepackage{supertabular}
\usepackage{placeins}
\usepackage{epsfig}
\usepackage{graphicx}
\usepackage{booktabs}
\usepackage{multirow}

\definecolor{purple}{rgb}{0.5,0,0.5}
\definecolor{blue}{rgb}{0.0,0,0.9}
\definecolor{prdblue}{rgb}{0.133,0.118,0.498}
\usepackage[colorlinks=true, pdfstartview=FitV, linkcolor=prdblue, citecolor= prdblue, urlcolor=prdblue]{hyperref}

\begin{document}

\title{Fully-charm tetraquarks: $cc\bar{c}\bar{c}$}
\author{Xiaoyun Chen}
\email[]{xychen@jit.edu.cn} \affiliation{College of Science, Jinling Institute of Technology, Nanjing 211169, P. R.
China}

\begin{abstract} \label{abstract}
In this work, we continue to study the mass spectra of fully-heavy tetraquarks $cc\bar{c}\bar{c}$ with the quantum numbers of $J^{PC}=0^{++}, 1^{+-}, 2^{++}$ in the nonrelativistic chiral quark model. With the help of the Gaussian Expansion Method, we present dynamical computations for $cc\bar{c}\bar{c}$ state with considering two structures, meson-meson [$\bar{c}c$][$\bar{c}c$] and diquark-antidiquark
[$cc$][$\bar{c}\bar{c}$] and their mixing. The results manifest that the energies of the low-lying states are all higher than the meson-meson thresholds [$\bar{c}c$][$\bar{c}c$]. However, resonances are possible because of the color structure. Several resonances are proposed and the lowest resonance is predicted to be 6.5 GeV and the stability of the resonance states is checked using the real scaling method.
\end{abstract}

\maketitle


\section{Introduction} \label{introduction}
In the past decades, a lot of charmonium-like/bottomonium-like $XYZ$ states~\cite{x3872,y4260-1,y4260-2,zc3900-1,zc3900-2,zc3900-3,zc3900-4,zb10610,pc} have been observed in experiment, which generates great challenges and opportunities for researchers to study the multiquark states.

Recently, the tetraquark of all-heavy system, such as $cc\bar{c}\bar{c}$ and $bb\bar{b}\bar{b}$ has received considerable attention due to the development of experiments. If the $cc\bar{c}\bar{c}$ or $bb\bar{b}\bar{b}$ state steadily exist, they are most likely to be observed at LHC and other facilities. In this work, we mainly concentrate on the $cc\bar{c}\bar{c}$ tetraquark state.

Whether there exist bound states of fully-charm tetraquarks has been debated for more than forty years, but there was no consensus until now. Theoretically, various methods are applied to study $cc\bar{c}\bar{c}$ states. 
In few works, it is suggested that there exists stable bound tetraquark $cc\bar{c}\bar{c}$ state~\cite{iwasaki,prd70014009,cpc43013105}.
Iwasaki~\cite{iwasaki} first argued that bound state of $c^2\bar{c}^2$ could exist and estimated its mass, which is in the neighborhood of 6 GeV or 6.2 GeV based on a string model.
Richard \emph{et al.} have used a parametrized Hamiltonian to calculate the spectrum of all-charm tetraquark state and found several close-lying bound states with two sets of parameters based on large but finite oscillator bases. For example, for the lowest state with quantum number $J^{PC}=0^{++}$, it had the mass below the threshold of two $\eta_c(1S)$, 5967.2 MeV~\cite{prd70014009}. In recent research, Debastiani \emph{et al.} used a non-relativistic model to study the spectroscopy of a tetraquark composed of $cc\bar{c}\bar{c}$ in a diquark-antidiquark configuration and found that the lowest $S$-wave $cc\bar{c}\bar{c}$ tetraquarks might be below their thresholds~\cite{cpc43013105}.

On the contrary, in some other works, there is no bound $cc\bar{c}\bar{c}$ state~\cite{prd73054004,prd86034004,prd95034011,prd97094015,plb773247,epjc77432,prd100016006,
prd252370}. Barnea \emph{et al.} studied the system consist of quarks and antiquarks of the same flavor within the hyperspherical formalism, and the mass of $cc\bar{c}\bar{c}$ is about 6038 MeV, which is above the corresponding threshold~\cite{prd73054004}. Karliner \emph{et al.} have calculated the mass spectrum of $cc\bar{c}\bar{c}$ state and found it unlikely to be less than twice mass of the lowest charmonium state $\eta_c$~\cite{prd95034011}. Recently in Ref.~\cite{prd100016006}, Ming-Sheng Liu \emph{et al.} suggested that no bound states could be formed below the thresholds of meson pairs $(c\bar{c})$-$(c\bar{c})$ within a potential model by including the linear confining potential, Coulomb potential and spin-spin interactions.

Hadron spectroscopy always played an important role in revealing the properties of the dynamics of strong interaction. In this paper, we investigate systematically the masses of $cc\bar{c}\bar{c}$ tetraquark state with $J^{PC}=0^{++}, 1^{+-}, 2^{++}$ in the quark model, which can describe well the properties of hadrons and hadron-hadron interactions. The method of Gasussian expansion method (GEM) is employed to do a high precision four-body
calculation. The dynamical mixing of the meson-meson configuration with the diquark-antidiquark configuration is also considered. All the color configurations,
color singlet-singlet $1\times1$ and color octet-octet $8\times8$ for meson-meson structure, and color antitriplet-triplet $\bar{3}\times3$ and sextet-antisextet $6\times\bar{6}$ for diquark-antidiquark structure, and their mixing are considered. This mixing occurs by both the spin-independent and the spin-dependent parts of the potential. To obtain the genuine resonances, the real scaling method (stabilization)~\cite{rsc1,rsc2} is applied in present work.

This paper is organized as follows. In Sec.~\ref{framework}, we briefly discuss the chiral quark model and the wave functions of $cc\bar{c}\bar{c}$, including the Gaussian Expansion Method. In Sec.~\ref{discussions}, the numerical results and discussion are presented. Some conclusions and summary are given in Sec.~\ref{epilogue}.

\section{Quark model and wave functions} \label{framework}
The chiral quark model has been successful both in describing the hadron spectra and hadron-hadron
interactions. The details of the model can be found in Refs.~\cite{094016chen,Vijande:2005}. For
$cc\bar{c}\bar{c}$ fully-heavy system, the Hamiltonian of the chiral quark
model consists of three parts: quark rest mass, kinetic energy, and potential energy:
\begin{align}
 H & = \sum_{i=1}^4 m_i  +\frac{p_{12}^2}{2\mu_{12}}+\frac{p_{34}^2}{2\mu_{34}}
  +\frac{p_{1234}^2}{2\mu_{1234}}  \quad  \nonumber \\
  & + \sum_{i<j=1}^4 \left( V_{ij}^{C}+V_{ij}^{G}\right).
\end{align}
The potential energy consists of pieces describing quark confinement (C) and one-gluon-exchange (G).
The detailed forms of potentials are shown below (only central parts are presented)
\cite{094016chen}: {\allowdisplaybreaks
\begin{subequations}
\begin{align}
V_{ij}^{C}&= ( -a_c r_{ij}^2-\Delta ) \boldsymbol{\lambda}_i^c
\cdot \boldsymbol{\lambda}_j^c ,  \\
 V_{ij}^{G}&= \frac{\alpha_s}{4} \boldsymbol{\lambda}_i^c \cdot \boldsymbol{\lambda}_{j}^c
\left[\frac{1}{r_{ij}}-\frac{2\pi}{3m_im_j}\boldsymbol{\sigma}_i\cdot
\boldsymbol{\sigma}_j
  \delta(\boldsymbol{r}_{ij})\right],  \\
\delta{(\boldsymbol{r}_{ij})} & =
\frac{e^{-r_{ij}/r_0(\mu_{ij})}}{4\pi r_{ij}r_0^2(\mu_{ij})}.
\end{align}
\end{subequations}}
$m_i$ is the constituent mass of quark/antiquark, and $\mu_{ij}$
is the reduced mass of two interacting quarks and
\begin{equation}
\mu_{1234}=\frac{(m_1+m_2)(m_3+m_4)}{m_1+m_2+m_3+m_4};
\end{equation}
$\mathbf{p}_{ij}=(\mathbf{p}_i-\mathbf{p}_j)/2$,
$\mathbf{p}_{1234}= (\mathbf{p}_{12}-\mathbf{p}_{34})/2$;
$r_0(\mu_{ij}) =s_0/\mu_{ij}$; $\boldsymbol{\sigma}$ are the $SU(2)$ Pauli matrices; $\boldsymbol{\lambda}$,
$\boldsymbol{\lambda}^c$ are $SU(3)$ flavor, color Gell-Mann matrices, respectively; and $\alpha_s$ is an
effective scale-dependent running coupling \cite{Vijande:2005},
\begin{equation}
\alpha_s(\mu_{ij})=\frac{\alpha_0}{\ln\left[(\mu_{ij}^2+\mu_0^2)/\Lambda_0^2\right]}.
\end{equation}
All the parameters are determined by fitting the meson spectrum, from light to heavy; and the resulting values are listed in Table~\ref{modelparameters}. Table~\ref{mesonmass} gives the theoretical masses of some charm mesons $c\bar{c}$ in the chiral quark model, also with the experimental data. Because of the orbital-spin interactions are not included in the calculation, the $P$-wave states $\chi_{cJ}, J=0,1,2$ have the same mass. 

\begin{table}[!t]
\begin{center}
\caption{ \label{modelparameters} Model parameters, determined by
fitting the meson spectrum from light to heavy.}
\begin{tabular}{llr}
\hline\hline\noalign{\smallskip}
Quark masses   &$m_u=m_d$    &313  \\
   (MeV)       &$m_s$         &536  \\
               &$m_c$         &1728 \\
               &$m_b$         &5112 \\
\hline
Confinement        &$a_c$ (MeV fm$^{-2}$)         &101 \\
                   &$\Delta$ (MeV)     &-78.3 \\
\hline
OGE                 & $\alpha_0$        &3.67 \\
                   &$\Lambda_0({\rm fm}^{-1})$ &0.033 \\
                  &$\mu_0$(MeV)    &36.98 \\
                   &$s_0$(MeV)    &28.17 \\
\hline\hline
\end{tabular}
\end{center}
\end{table}

\begin{table}[!t]
\begin{center}
\caption{ \label{mesonmass} The masses of some heavy mesons (unit: MeV). $M_{cal}$ and $M_{exp}$ represent the theoretical
and the experimental masses, respectively.}
\begin{tabular}{ccccccc}
\hline\hline\noalign{\smallskip}
meson     & $\eta_c$  & $J/\psi$ & $h_{c}$ & $\chi_{c0}$ & $\chi_{c1}$ & $\chi_{c2}$  \\
\hline
$M_{cal}$ &   2986.3    & 3096.4   &3417.3    &3416.4      &3416.4   &3416.4         \\
$M_{exp}$ &   $2983.4$    & $3096.9$   & $3525.38$    &  $3414.75$
   & $3510.66$  & $3556.20$  \\
\hline\hline
\end{tabular}
\end{center}
\end{table}

The wave functions of four-quark states for the two structures, diquark-antidiquark and
meson-meson, can be constructed in two steps. For each degree of freedom, first we construct
the wave functions for two-body sub-clusters, then couple the wave functions of two sub-clusters to obtain the wave functions of four-quark states.

(1) Diquark-antidiquark structure.

For the spin part, the wave functions for two-body sub-clusters are,
\begin{align}
&\chi_{11}=\alpha\alpha,~~
\chi_{10}=\frac{1}{\sqrt{2}}(\alpha\beta+\beta\alpha),~~
\chi_{1-1}=\beta\beta,\nonumber \\
&\chi_{00}=\frac{1}{\sqrt{2}}(\alpha\beta-\beta\alpha),
\end{align}
then the wave functions for four-quark states are obtained,
 {\allowdisplaybreaks
\begin{subequations}\label{spinwavefunctions}
\begin{align}
\chi_{00}^{\sigma
1}&=\chi_{00}\chi_{00},\\
\chi_{00}^{\sigma
2}&=\sqrt{\frac{1}{3}}(\chi_{11}\chi_{1-1}-\chi_{10}\chi_{10}+\chi_{1-1}\chi_{11}),\\
\chi_{11}^{\sigma
3}&=\chi_{00}\chi_{11},\\
 \chi_{11}^{\sigma
4}&=\chi_{11}\chi_{00},\\
\chi_{11}^{\sigma
5}&=\frac{1}{\sqrt{2}}(\chi_{11}\chi_{10}-\chi_{10}\chi_{11}),\\
\chi_{22}^{\sigma 6}&=\chi_{11}\chi_{11}.
\end{align}
\end{subequations}}
Where the superscript $\sigma i$ $(i=1 \sim 6)$ of $\chi$
represents the index of the spin wave functions of four-quark
states. The subscripts of $\chi$ are $SM_S$, the total spin and
the third projection of total spin of the system. $S=0, 1, 2$, and
only one component ($M_S=S$) is shown for a given total spin $S$.

The wave function for the flavor part is very simple,
\begin{equation}
\chi_{d0}^{f} = (cc)(\bar{c}\bar{c}).
\end{equation}
The subscript $d0$ of $\chi$ represents the diquark-antidiquark structure and isospin ($I=0$).

For the color part, the wave functions of four-quark states must be color singlet $[222]$
and it is obtained as below,
\begin{subequations}
\begin{align}
\chi^{c1}_{d} & =
\frac{\sqrt{3}}{6}(rg\bar{r}\bar{g}-rg\bar{g}\bar{r}+gr\bar{g}\bar{r}-gr\bar{r}\bar{g} \nonumber \\
&~~~+rb\bar{r}\bar{b}-rb\bar{b}\bar{r}+br\bar{b}\bar{r}-br\bar{r}\bar{b} \nonumber \\
&~~~+gb\bar{g}\bar{b}-gb\bar{b}\bar{g}+bg\bar{b}\bar{g}-bg\bar{g}\bar{b}).  \\
\chi^{c2}_{d}&=\frac{\sqrt{6}}{12}(2rr\bar{r}\bar{r}+2gg\bar{g}\bar{g}+2bb\bar{b}\bar{b}
    +rg\bar{r}\bar{g}+rg\bar{g}\bar{r} \nonumber \\
&~~~+gr\bar{g}\bar{r}+gr\bar{r}\bar{g}+rb\bar{r}\bar{b}+rb\bar{b}\bar{r}+br\bar{b}\bar{r} \nonumber \\
&~~~+br\bar{r}\bar{b}+gb\bar{g}\bar{b}+gb\bar{b}\bar{g}+bg\bar{b}\bar{g}+bg\bar{g}\bar{b}).
\end{align}
\end{subequations}
Where, $\chi_{d}^{c1}$ and $\chi_{d}^{c2}$ represents the color antitriplet-triplet
($\bar{3}\times3$) and sextet-antisextet ($6\times\bar{6}$) coupling, respectively. The detailed
coupling process for the color wave functions can refer to our previous work \cite{054022chen}.

(2) Meson-meson structure.

For the spin part, the wave functions are the same as those of the diquark-antidiquark structure,
Eq.~(\ref{spinwavefunctions}).

For the flavor part, the wave function is,
\begin{equation}
\chi_{m0}^{f}=(\bar{c}c)(\bar{c}c),
\end{equation}
The subscript $m0$ of $\chi$ represents the meson-meson structure
and isospin equals zero.

For the color part, the wave functions of four-quark states in the meson-meson structure are,
\begin{subequations}
\begin{align}
\chi_{m}^{c1}&=\frac{1}{3}(\bar{r}r+\bar{g}g+\bar{b}b)(\bar{r}r+\bar{g}g+\bar{b}b),\\
\chi_{m}^{c2}&=\frac{\sqrt{2}}{12}(3\bar{b}r\bar{r}b+3\bar{g}r\bar{r}g+3\bar{b}g\bar{g}b+3\bar{g}b\bar{b}g+3\bar{r}g\bar{g}r \nonumber \\
&~~~+3\bar{r}b\bar{b}r+2\bar{r}r\bar{r}r+2\bar{g}g\bar{g}g+2\bar{b}b\bar{b}b-\bar{r}r\bar{g}g \nonumber\\
&~~~-\bar{g}g\bar{r}r-\bar{b}b\bar{g}g-\bar{b}b\bar{r}r-\bar{g}g\bar{b}b-\bar{r}r\bar{b}b).
\end{align}
\end{subequations}
Where, $\chi_{m}^{c1}$ and $\chi_{m}^{c2}$ represents the color singlet-singlet ($1\times1$) and
color octet-octet ($8\times8$) coupling, respectively. The details refer to our previous work
\cite{054022chen}.

As for the orbital wave functions, they can be constructed by coupling the orbital wave function
for each relative motion of the system,
\begin{equation}\label{spatialwavefunctions}
\Psi_{L}^{M_{L}}=\left[[\Psi_{l_1}({\bf r}_{12})\Psi_{l_2}({\bf
r}_{34})]_{l_{12}}\Psi_{L_r}({\bf r}_{1234}) \right]_{L}^{M_{L}},
\end{equation}
where $l_1$ and $l_2$ is the angular momentum of two sub-clusters, respectively.
$\Psi_{L_r}(\mathbf{r}_{1234})$ is the wave function of the relative motion between two
sub-clusters with orbital angular momentum $L_r$. $L$ is the total orbital angular momentum
of four-quark states. Here for the low-lying $cc\bar{c}\bar{c}$ state,
all angular momentum ($l_1, l_2, L_r, L$) are taken as zero. The used Jacobi coordinates
are defined as,
\begin{align}\label{jacobi}
{\bf r}_{12}&={\bf r}_1-{\bf r}_2, \nonumber \\
{\bf r}_{34}&={\bf r}_3-{\bf r}_4, \nonumber\\
{\bf r}_{1234}&=\frac{m_1{\bf r}_1+m_2{\bf
r}_2}{m_1+m_2}-\frac{m_3{\bf r}_3+m_4{\bf r}_4}{m_3+m_4}.
\end{align}
For diquark-antidiquark structure, the quarks are numbered as $1, 2$, and the antiquarks are
numbered as $3, 4$; for meson-meson structure, the antiquark and quark in one cluster are
marked as $1, 2$, the other antiquark and quark are marked as $3, 4$. In the two structure
coupling calculation, the indices of quarks, antiquarks in diquark-antidiquark structure will be
changed to be consistent with the numbering scheme in meson-meson structure.
In GEM, the spatial wave function is expanded by Gaussians~\cite{GEM}:
\begin{subequations}
\label{radialpart}
\begin{align}
\Psi_{l}^{m}(\mathbf{r}) & = \sum_{n=1}^{n_{\rm max}} c_{n}\psi^G_{nlm}(\mathbf{r}),\\
\psi^G_{nlm}(\mathbf{r}) & = N_{nl}r^{l}
e^{-\nu_{n}r^2}Y_{lm}(\hat{\mathbf{r}}),
\end{align}
\end{subequations}
where $N_{nl}$ are normalization constants,
\begin{align}
N_{nl}=\left[\frac{2^{l+2}(2\nu_{n})^{l+\frac{3}{2}}}{\sqrt{\pi}(2l+1)}
\right]^\frac{1}{2}.
\end{align}
$c_n$ are the variational parameters, which are determined dynamically. The Gaussian size
parameters are chosen according to the following geometric progression
\begin{equation}\label{gaussiansize}
\nu_{n}=\frac{1}{r^2_n}, \quad r_n=r_1a^{n-1}, \quad
a=\left(\frac{r_{n_{\rm max}}}{r_1}\right)^{\frac{1}{n_{\rm
max}-1}}.
\end{equation}
This procedure enables optimization of the expansion using just a small numbers of Gaussians.
Finally, the complete channel wave function for the four-quark system for diquark-antidiquark
structure is written as
\begin{align}\label{diquarkpsi}
&\Psi_{IJ,i,j}^{M_IM_J}={\cal
A}_1[\Psi_{L}^{M_{L}}\chi_{SM_{S}}^{\sigma
i}]_{J}^{M_J}\chi_{d0}^{f}\chi^{cj}_{d},
  \nonumber \\
&(i=1\sim6; j=1,2; S=0,1,2),
\end{align}
where ${\cal A}_1$ is the antisymmetrization operator, for
$cc\bar{c}\bar{c}$ system,
\begin{equation}
{\cal A}_1=\frac{1}{2}(1-P_{12}-P_{34}+P_{12}P_{34}).
\end{equation}

For meson-meson structure, the complete wave function is written as
\begin{align}
&\Psi_{IJ,i,j}^{M_IM_J}= {\cal
A}_2[\Psi_{L}^{M_{L}}\chi_{SM_{S}}^{\sigma
i}]_{J}^{M_J}\chi_{m0}^{f}\chi^{cj}_{m},\nonumber \label{mesonpsi}\\
&(i=1\sim6; j=1,2; S=0,1,2),
\end{align}
where ${\cal A}_2$ is the antisymmetrization operator, for
$c\bar{c}c\bar{c}$ system,
\begin{equation}
{\cal A}_2=\frac{1}{2}(1-P_{13}-P_{24}+P_{13}P_{24}).
\end{equation}

Lastly, the eigenenergies of the four-quark system are obtained by solving a
Schr\"{o}dinger equation:
\begin{equation}
    H \, \Psi^{\,M_IM_J}_{IJ}=E^{IJ} \Psi^{\,M_IM_J}_{IJ},
\end{equation}
where $\Psi^{\,M_IM_J}_{IJ}$ is the wave function of the four-quark states, which is the
linear combinations of the above channel wave functions, Eq.~(\ref{diquarkpsi}) in the
diquark-antidiquark structure or Eq.~(\ref{mesonpsi}) in the meson-meson structure, or
both wave functions of Eq.~(\ref{diquarkpsi}) and (\ref{mesonpsi}), respectively.

\section{Results and discussions} \label{discussions}
In this work, we estimated the masses of the lowest-lying $cc\bar{c}\bar{c}$ tetraquark state with quantum numbers $J^{PC}=0^{++}, 1^{+-}, 2^{++}$ in the chiral quark model by adopting GEM. The pure meson-meson and the pure diquark-antidiquark structure, along with the dynamical mixing of these two structures are considered, respectively.
In our calculations, all possible color, and spin configurations are included. For example, for meson-meson structure, two color configurations, color singlet-singlet ($1 \times 1$) and octet-octet ($8 \times 8$) are employed; for diquark-antidiquark structure, color antitriplet-triplet ($\bar{3} \times 3$) and sextet-antisextet ($6 \times \bar{6}$) are taken into account. In Table ~\ref{index}, we give the index of channel wave functions. The Pauli principle forbidden channels have been elimanited. For $J^{PC}=0^{++}$, there are six channels, four in meson-meson structure
and two in diquark-antidiquark structure. For $J^{PC}=1^{++}$, there are five channels,
four in meson-meson structure and only one in diquark-antidiquark structure. For $J^{PC}=2^{++}$, the total channels are three, with 2 in meson-meson structure
and one in diquark-antidiquark structure.

\begin{table}[!t]
\begin{center}
\caption{ \label{index} The index of channel wave functions. }
\begin{tabular}{cccccc} \hline \hline
\multicolumn{2}{c}{$J^{PC}=0^{++}$} & \multicolumn{2}{c}{$J^{PC}=1^{+-}$} &
 \multicolumn{2}{c}{$J^{PC}=2^{++}$} \\  \hline
 1  & ~~$\chi_{00}^{\sigma 1}\chi_{m0}^f\chi_{m}^{c1}$ ~~ &
 1  & ~~$\chi_{11}^{\sigma 3}\chi_{m0}^f\chi_{m}^{c1}$ ~~ &
 1  & $\chi_{22}^{\sigma 6}\chi_{m0}^f\chi_{m}^{c1}$  \\
 2  & $\chi_{00}^{\sigma 1}\chi_{m0}^f\chi_{m}^{c2}$  &
 2  & $\chi_{11}^{\sigma 3}\chi_{m0}^f\chi_{m}^{c2}$  &
 2  & $\chi_{22}^{\sigma 6}\chi_{m0}^f\chi_{m}^{c2}$  \\
 3  & $\chi_{00}^{\sigma 2}\chi_{m0}^f\chi_{m}^{c1}$  &
 3  & $\chi_{11}^{\sigma 4}\chi_{m0}^f\chi_{m}^{c1}$  &
 3  & $\chi_{22}^{\sigma 6}\chi_{d0}^f\chi_{d}^{c1}$  \\
 4  & $\chi_{00}^{\sigma 2}\chi_{m0}^f\chi_{m}^{c2}$  &
 4  & $\chi_{11}^{\sigma 4}\chi_{m0}^f\chi_{m}^{c2}$  & & \\
 5  & $\chi_{00}^{\sigma 2}\chi_{d0}^f\chi_{d}^{c1}$  &
 5  & $\chi_{11}^{\sigma 5}\chi_{d0}^f\chi_{d}^{c1}$  & & \\
 6  & $\chi_{00}^{\sigma 1}\chi_{d0}^f\chi_{d}^{c2}$  & & & & \\
\hline \hline
\end{tabular}
\end{center}
\end{table}

\begin{table}[!t]
\begin{center}
\caption{ \label{results1} The results of $cc\bar{c}\bar{c}$ states
with $J^{PC}=0^{++}$ in pure meson-meson structure, diquark-antidiquark structure, and
in considering the mixing of two structures, respectively.
"$E_{th}^{theo}$" represents the theoretical thresholds. (unit: MeV).}
\begin{tabular}{cccc} \hline \hline
Channel     & $E$      & $E_{th}^{theo}$ & $E_{th}^{exp}$ \\ \hline
 1        & ~~5973.4~~  & ~~5972.6~~         & ~~5966.8 ~~  \\
 2        & 6373.2  &                 &          \\
 3        & 6193.7  & 6192.8         & 6193.8   \\
 4        & 6356.9  &                 &          \\
 5        & 6360.2  &                 &          \\
 6        & 6390.9  &                 &          \\
 1+2+3+4  & 5973.4  & 5972.6         & 5966.8   \\
 5+6      & 6345.7  &                 &          \\
 1+2+3+4+5+6  & 5973.4  & 5972.6     & 5966.8   \\  \hline \hline
\end{tabular}
\end{center}
\end{table}

\begin{table}[!t]
\begin{center}
\caption{ \label{results2} The results of $cc\bar{c}\bar{c}$ states
with $J^{PC}=1^{+-}$ in pure meson-meson structure, diquark-antidiquark structure, and
in considering the mixing of two structures, respectively.
"$E_{th}^{theo}$" represents the theoretical thresholds. (unit: MeV).}
\begin{tabular}{ccccc} \hline \hline
Channel      & $E$      & $E_{th}^{theo}$ & $E_{th}^{exp}$ \\ \hline \hline
 1         & ~~6083.6~~  & ~~6082.7~~ & ~~6080.3~~   \\
 2         & 6349.8  &         &          \\
 3         & 6083.6  & 6082.7 & 6080.3   \\
 4         & 6349.8  &         &          \\
 5         & 6397.6  &         &          \\
 1+2       & 6083.6  & 6082.7 & 6080.3   \\
 1+2+3+4+5 & 6083.6  & 6082.7 & 6080.3   \\
\hline \hline
\end{tabular}
\end{center}
\end{table}

\begin{table}[!t]
\begin{center}
\caption{ \label{results3} The results of $cc\bar{c}\bar{c}$ states
with $J^{PC}=2^{++}$ in pure meson-meson structure, diquark-antidiquark structure, and
in considering the mixing of two structures, respectively.
"$E_{th}^{theo}$" represents the theoretical thresholds. (unit: MeV).}
\begin{tabular}{ccccc} \hline \hline
Channel      & $E$      & $E_{th}^{theo}$ & $E_{th}^{exp}$ \\ \hline \hline
 1         & ~~6193.7~~  & ~~6192.8~~  & ~~6193.8~~  \\
 2         & 6365.3  &          &         \\
 3         & 6410.4  &          &         \\
 1+2       & 6193.7  & 6192.8  & 6193.8  \\
 1+2+3     & 6193.7  & 6192.8  & 6193.8  \\ \hline \hline
\end{tabular}
\end{center}
\end{table}

The single-channel and channel-coupling calculations are performed in the present work. Tables ~\ref{results1}-\ref{results3} give the results of $cc\bar{c}\bar{c}$ tetraquarks with quantum numbers $J^{PC}=0^{++}, 1^{+-}, 2^{++}$, respectively. From tables, we found that
the coupling of the color configurations $1 \times 1$ and $8 \times 8$ in meson-meson structure is rather small, but the coupling of the color configurations $\bar{3} \times 3$ and $6 \times \bar{6}$ plays a role in diquark-antidiquark structure.
And the energies in diquark-antidiquark structure are all much larger than those in meson-meson structure. After considering the mixing of two structures, we found that the effects of the two-structure mixing seem to be tiny for the lowest-lying energies and 
finally the ground state energies $E$ for $0^{++}, 1^{+-}, 2^{++}$ in each case are shown
in the second and third column of Tables ~\ref{results1}-\ref{results3} respectively, which are all a litter higher than the corresponding theoretical thresholds which are
given in the last column of the tables. No bound states are formed in our calculations for $cc\bar{c}\bar{c}$ tetraquarks.

\begin{figure}
\centering
\includegraphics[width=0.5\textwidth]{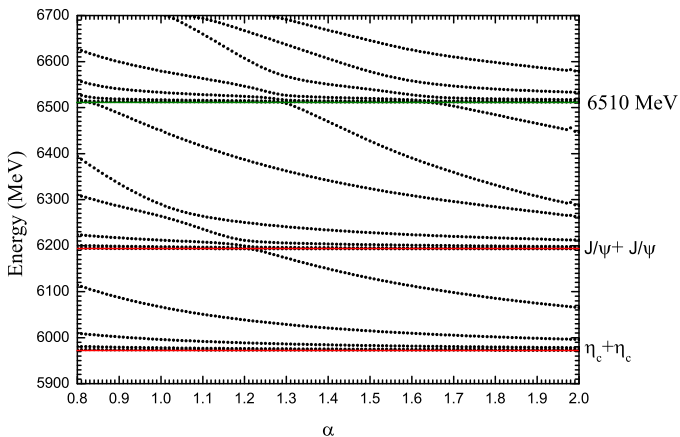}
\caption{\label{rsc00} The stabilization plots of the energies of
$cc\bar{c}\bar{c}$ states for $J^{PC}=0^{++}$ with the respect to the scaling factor $\alpha$.}
\end{figure}

\begin{figure}
\centering
\includegraphics[width=0.5\textwidth]{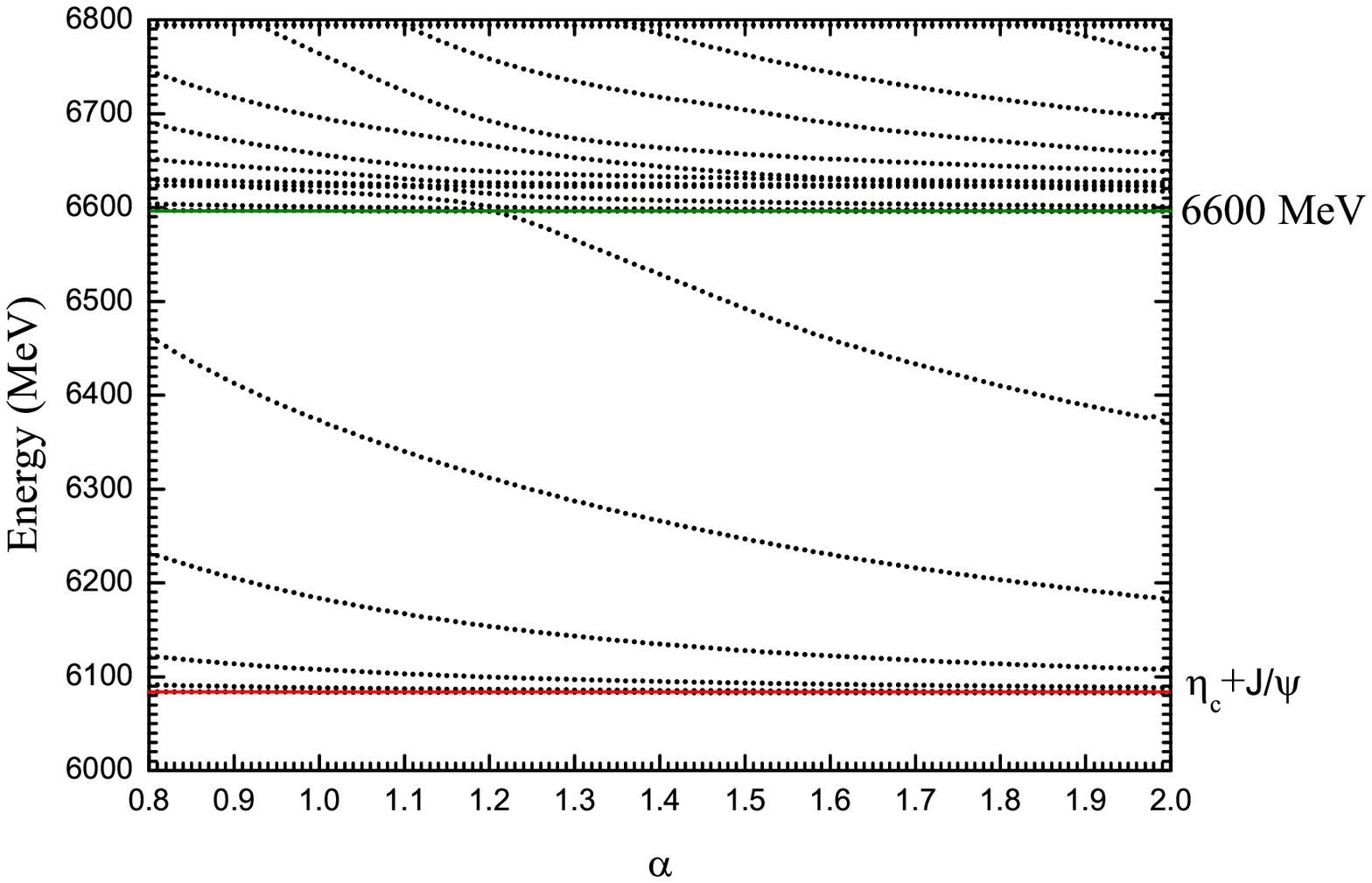}
\caption{\label{rsc01} The stabilization plots of the energies of
$cc\bar{c}\bar{c}$ states for $J^{PC}=1^{+-}$ with the respect to the scaling factor $\alpha$.}
\end{figure}

\begin{figure}
\centering
\includegraphics[width=0.5\textwidth]{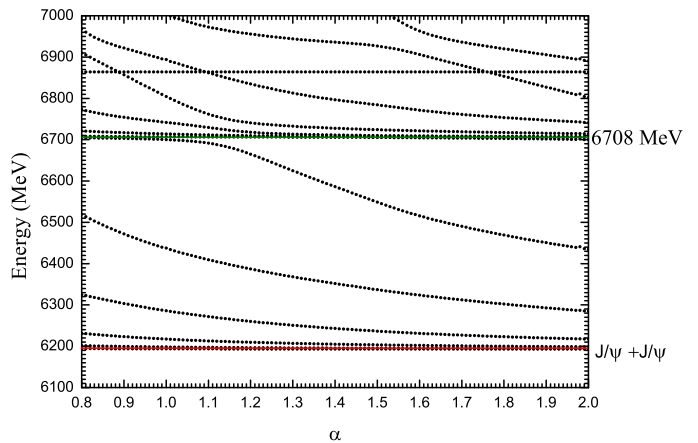}
\caption{\label{rsc02} The stabilization plots of the energies of
$cc\bar{c}\bar{c}$ states for $J^{PC}=2^{++}$ with the respect to the scaling factor $\alpha$.}
\end{figure}

Even with the higher energies than the thresholds of $cc\bar{c}\bar{c}$ tetraquark, there possibly exists resonances because the color structures of the system. In present work, we employ the dedicated real scaling (stabilization) method and try to find the genuine resonances. The real scaling method was often used for analyzing electron-atom and electron-molecule scattering~\cite{RSM}. In the present approach, the real scaling method is realized by scaling the Gaussian size parameters $r_n$ in Eq.~\ref{gaussiansize} just for the meson-meson structure with the $1 \times 1$ color configuration, i.e., $r_n \rightarrow \alpha r_n$, where $\alpha$ takes the values between 0.8 and 2.0. We illustrate the energies for $cc\bar{c}\bar{c}$ tetraquarks for $J^{PC}=0^{++}, 1^{+-}, 2^{++}$ with the respect to the scaling factor $\alpha$ in Figs~\ref{rsc00}-~\ref{rsc02}, respectively. From the figures, we can clearly see that with the increasing $\alpha$, most of the states fall off towards their thresholds, but there are several states with stable energies, and they are thresholds or the genuine resonances. Thresholds are marked with
the physical contents, for example, $\eta_c+\eta_c$ and $J/\psi+J/\psi$ in Fig.~1, the genuine resonances are marked by their energies, for instance, 6510 MeV in Fig.~1.

For $0^{++}$ states (Fig.~\ref{rsc00}), there are three stable energies under 6700 MeV. The first two stable energies represent the two thresholds $2\eta_c$ (spin $0 \otimes 0 \rightarrow 0$) and $2J/ \Psi$ (spin $1 \otimes 1 \rightarrow 0$). The third stable energy around 6510 MeV is exactly the genuine resonance what we are looking for and its energy is stable against the variation of the scale factor $\alpha$. For $1^{+-}$ and $2^{++}$ states in Fig.~\ref{rsc01} and Fig.~\ref{rsc02}, we found that the genuine resonance is about 6600 MeV and 6708 MeV, respectively.

In order to identify the structures of these possible resonances, we calculate the distance between $c$ and $\bar{c}$ quark, denoted as $R_{c\bar{c}}$, as well as the distance between $c$ and $c$ quark, denoted as $R_{cc}$ for the resonance states, respectively, which are shown in Table~\ref{distance}. From the table, we can see that $R_{cc}$ is rather large, it means that the state is very likely to be molecular one.
the large $R_{c\bar{c}}$ is due to the antisymmetrization, it gives the average distance
between $c$ and two $\bar{c}$. The distance $R^{\prime}_{c\bar{c}}$ between $c$ and $\bar{c}$ in one sub-cluster can be extracted from $R_{cc}$ and $R_{c\bar{c}}$, which is
shown in the last column of Table~\ref{distance}. From $R^{\prime}_{cc}$ and $R_{c\bar{c}}$, we can see that the three resonances listed in Table~\ref{distance} are
molecules.

For comparison, in Table~\ref{comparison}, our predicted resonance masses and other estimation of $cc\bar{c}\bar{c}$ tetraquark are summarized. It shows that our predicted masses for $cc\bar{c}\bar{c}$ tetraquark are roughly consistent with the nonrelativistic quark model predictions of Refs~\cite{prd70014009,prd97094015,prd100016006} and results obtained by QCD sum rules in Ref~\cite{plb773247}. But for other results in Table~\ref{comparison}, the masses are all lower than our predictions. the reason may be that the simplified interaction between quarks are used or a restrictive structure diquark-antidiquark picture is applied. All these masses give mixed signals, more experimental information from the Belle-II and LHCb analyses would be able to clarify these issues in the near future.

\begin{table}[!t]
\begin{center}
\caption{ \label{distance} The distances between $c$ and $c(\bar{c})$ quark for the possible resonance states of $cc\bar{c}\bar{c}$ system. $R^{\prime}_{c\bar{c}}$ denotes
the distance between $c$ and $\bar{c}$ in the subcluster.}
\begin{tabular}{ccccc} \hline \hline
State             & Resonance (MeV)      & $R_{c\bar{c}}$ (fm) & $R_{cc}$ (fm) & $R^{\prime}_{c\bar{c}}$ (fm) \\ \hline 
 $0^{++}$         & ~~6510~~  & ~~3.41~~  & ~~4.78~~   & 0.6  \\
 $1^{+-}$         & 6600      &   2.63    &   3.67     & 0.6  \\
 $2^{++}$         & 6708      &   2.86    &   3.98     & 0.7  \\ \hline \hline
\end{tabular}
\end{center}
\end{table}

\begin{table*}[!t]
	\begin{center}
		\caption{ \label{comparison} Predictions for the masses of the $cc\bar{c}\bar{c}$ tetraquark.}
		\begin{tabular}{ccccccccccccccc} \hline \hline
			State        &This work   &\cite{prd86034004}  &\cite{prd100016006} &\cite{prd97094015} &\cite{prd70014009} &\cite{epjc77432} &\cite{cpc43013105} &\cite{plb773247} &\cite{prd73054004} &\cite{prd252370} &\cite{prd95034011} &\cite{plb718545} &\cite{iwasaki} &\cite{epjc78647} \\
			$0^{++}$    & 6510   &5966   &6487  &6797 &6477 &5990 &5969 &6460$\sim$6470 &6038$\sim$6115   &6383 &6192$\pm$25  &5300$\pm$500  &$\sim$6200   & $<$6140  \\
			$1^{+-}$    & 6600   &6051   &6500  &6899 &6528 &6050 &6021 &6370$\sim$6510 &6101$\sim$6176   &6437 &...          &...            &...         &...    \\
			$2^{++}$    & 6708   &6223   &6524  &6956 &6573 &6090 &6115 &6370$\sim$6510 &6172$\sim$6216   &6437 &...          &...            &...         &...    \\ \hline\hline
		\end{tabular}
	\end{center}
\end{table*}

\section{Summary} \label{epilogue}
In this work, we study the mass spectra of the fully-charm $cc\bar{c}\bar{c}$ system with quantum numbers $J^{PC}=0^{++}, 1^{+-}, 2^{++}$ in the chiral quark model with the help of GEM. The dynamical mixing of the meson-meson structure and the diquark-antidiquark structure, along with all possible color, spin configurations are taken into account. The predicted masses of the lowest-lying $cc\bar{c}\bar{c}$ states are all above the corresponding two meson decay thresholds, leaving no space for bound states. By adopting the real scaling method, it suggests that there exist possible lowest resonances for $J^{PC}=0^{++}, 1^{+-}, 2^{++}$ states, with masses 6510 MeV, 6600 MeV and 6708 MeV, respectively. 

In general, the $QQ\bar{Q}\bar{Q}$ ($Q = b, c$) resonance states mainly decay into two $Q\bar{Q}$ meson final state by spontaneous dissociation. For the fully-charm $cc\bar{c}\bar{c}$ tetraquarks, they can decay via the spontaneous dissociation mechanism since they lie above the two-charmonium thresholds.
Because of the much heavier energies than the conventional charmonium mesons $c\bar{c}$, the doubly hidden-charm tetraquarks can be clearly differentiated in experiment. But it is more difficult for the production of the $cc\bar{c}\bar{c}$ states because two heavy quark pairs need to be created in the vacuum. However, the recent observations of the $J/ \psi J/ \psi$~\cite{plb70752,jp09094}, $J/ \psi \Upsilon(1S)$~\cite{prl116082002} and $\Upsilon(1S)\Upsilon(1S)$~\cite{jp05013} events bring some hope for the production of $cc\bar{c}\bar{c}$ tetraquarks. So it is a good choice to search for the $cc\bar{c}\bar{c}$ tetraquarks in the $J/ \psi J/ \psi$ and $\eta_c(1S)\eta_c(1S)$ channels. Our research provide
some useful information about the $cc\bar{c}\bar{c}$ tetraquark. In the near future, it is hopeful that $cc\bar{c}\bar{c}$ tetraquark can be observed in experiment.

\acknowledgments This work is supported partly by the National Natural
Science Foundation of China under Contract Nos. 11847145 and 11775118.


\end{document}